\documentclass[a4paper,aps,prl,preprintnumbers,twocolumn,amsmath,amssymb]{revtex4}

\usepackage{amsmath,amsfonts,amssymb,graphics,graphicx,epsfig,color,times,bbm}

\begin{document}
\bibliographystyle{apsrev}

\newcommand{\R}{\mathbbm{R}}
\newcommand{\rr}{\mathbbm{R}}
\newcommand{\nn}{\mathbbm{N}}
\newcommand{\cc}{\mathbbm{C}}
\newcommand{\ii}{\mathbbm{1}}
\newcommand{\M}{{\cal M}}
\newcommand{\T}{{\cal T}}
\newcommand{\1}{\mathbbm{1}}
\newcommand{\id}{{\rm id}}

\newcommand{\tr}{{\rm tr}}
\newcommand{\gr}[1]{\boldsymbol{#1}}
\newcommand{\be}{\begin{equation}}
\newcommand{\ee}{\end{equation}}
\newcommand{\bea}{\begin{eqnarray}}
\newcommand{\eea}{\end{eqnarray}}
\newcommand{\ket}[1]{|#1\rangle}
\newcommand{\bra}[1]{\langle#1|}
\newcommand{\avr}[1]{\langle#1\rangle}
\newcommand{\G}{{\cal G}}
\newcommand{\eq}[1]{Eq.~(\ref{#1})}
\newcommand{\ineq}[1]{Ineq.~(\ref{#1})}
\newcommand{\sirsection}[1]{\section{\large \sf \textbf{#1}}}
\newcommand{\sirsubsection}[1]{\subsection{\normalsize \sf \textbf{#1}}}
\newcommand{\ack}{\subsection*{\normalsize \sf \textbf{Acknowledgements}}}
\newcommand{\front}[5]{\title{\sf \textbf{\Large #1}}
\author{#2 \vspace*{.4cm}\\
\footnotesize #3}
\date{\footnotesize \sf \begin{quote}
\hspace*{.2cm}#4 \end{quote}
#5} \maketitle}
\newcommand{\eg}{\emph{e.g.}~}

\newcommand{\proofend}{\hfill\fbox\\\medskip }


\newtheorem{theorem}{Theorem}
\newtheorem{proposition}{Proposition}

\newtheorem{lemma}{Lemma}

\newtheorem{definition}{Definition}
\newtheorem{corollary}{Corollary}
\newtheorem{example}{Example}

\newcommand{\proof}[1]{{\bf Proof.} #1 $\proofend$}

\title{A not-so-normal mode decomposition}

\author{Michael M. Wolf}
\affiliation{ Max-Planck-Institute for Quantum Optics,
 Hans-Kopfermann-Str.\ 1, D-85748 Garching, Germany.}

\date{\today}


\begin{abstract} We provide a generalization of the normal mode
decomposition for non-symmetric or locality constrained
situations. This allows for instance to locally decouple a
bipartitioned collection of arbitrarily correlated oscillators up
to elementary pairs into which all correlations are condensed.
Similarly, it enables us to decouple the interaction parts of
multi-mode channels into single-mode and pair-interactions where
the latter are shown to be a clear signature of squeezing between
system and environment. In mathematical terms the result is a
canonical matrix form with respect to real symplectic equivalence
transformations.
\end{abstract}

\maketitle

The normal mode decomposition is a ubiquitous and indispensable
tool in physics and engineering. It allows us to transform into
frames where seemingly complex systems decouple into elementary
units each of which can be tackled individually. Consider for
instance a collection of harmonically coupled oscillators,
classical or quantum. Then the normal mode decomposition provides
us with a canonical transformation which decouples the state of
all oscillators into independent normal modes. Similarly, by the
seminal work of Williamson \cite{Williamson}, the evolution of a
system governed by a quadratic Hamiltonian can be decoupled into
independent elementary parts. In both cases---states and
evolutions---this decoupling requires a global transformation. In
many situations, however, additional restrictions prevent us from
simplifying the structure of a system by such a transformation. We
might for instance want to preserve the amount of correlations or
entanglement of a bipartitioned collection of oscillators. Or, in
the case of evolution, we might be faced with dissipation, i.e.,
an inaccessible environment.

In this work we provide a generalization of the normal mode
decomposition which is applicable in these and other situations
and allows again to decouple correlations and interactions into
elementary parts. This will reveal a remarkable structure of both,
correlations and interactions, in which elementary units turn out
to be pairs of modes rather than single modes.
Hence, as for the normal mode decomposition, we get again a
significant simplification in many contexts dealing with many-body
harmonic or bosonic systems, albeit with a richer structure.

In the first part we will state and prove the main result in terms
of matrix analysis where it amounts to a normal form with respect
to symplectic equivalence transformations, i.e., a symplectic
analogue of the singular value decomposition---inspired by recent
advances in symplectic geometry \cite{root,cSVD}. In the second
part we will then apply it to the above mentioned cases of
bipartite correlations and dissipative evolutions (depicted in
Figs. 1,2) and connect it to known results.

\section{Preliminaries}

Before introducing some basic notions let us mention that although
we will have quantum systems in mind in the following, all results
hold for classical systems in exactly the same way. Similarly,
note that the oscillators do not have to be mechanical but might
as well correspond to electromagnetical field modes, charge-phase
oscillations in Josephson junctions or collective spin
fluctuations, e.g., in atomic ensembles. In each of these cases we
have pairs of canonically conjugate variables obeying the same
(commutation) relation as position and momentum.

Consider now $n$ quantum mechanical oscillators characterized by a
set of momentum and position operators
$(P_1,\ldots,P_n,Q_1,\ldots,Q_n)=:R$ which obey the canonical
commutation relations
$[R_k,R_l]=i\sigma_{kl}$, with \be \sigma=\left(%
\begin{array}{cc}
  0 & -\1_n \\
  \1_n & 0 \\
\end{array}%
\right)\ee the symplectic matrix. A canonical/symplectic
transformation maps $R_k\mapsto \sum_l S_{kl}R_k$ such that the
commutation relations (or classical Poisson brackets) are
preserved, i.e., $S\sigma S^T=\sigma$. We will denote the group of
real symplectic transformation on $n$ modes by $Sp(2n)$. For a
basic introduction into symplectic transformations and the
appearance and use of canonically conjugate variables in quantum
information theory we refer the reader to \cite{Mukunda} and
\cite{BL}.

The essence of the ordinary normal mode decomposition
\cite{Williamson} is the fact that for any positive definite
matrix $X\in\mathbb{R}^{2n\times 2n}$ there is an $S\in Sp(2n)$
such that \be \label{eq:Williamson}SXS^T={\rm
diag}(\nu_1,\ldots,\nu_n,\nu_1,\ldots\nu_n).\ee In case $X$
represents a Hamiltonian $H=\sum_{kl}X_{kl}R_k R_l$ the $\nu_k$
are the normal mode frequencies. If $X_{kl}=\langle\{R_k-\langle
R_k\rangle,R_l-\langle R_l\rangle\}_+\rangle$ is a covariance
matrix, then $(\nu_k-1)/2$ is the mean occupation number
(phonons/photons) in the $k$'th normal mode.

\section{Canonical form}
We aim at deriving a canonical form for general (not necessarily
symmetric)  matrices $X\in\mathbb{R}^{2n\times 2n}$ under
symplectic equivalence transformations $X\mapsto S_1X S_2$. To
this end we will first construct a set of invariants which play a
role similar to the normal mode frequencies $\nu_k$:

\begin{proposition}[Invariants]\label{prop:inv}
The eigenvalues of $\Sigma(X):=X\sigma X^T\sigma^T$ are invariant
with respect to symplectic equivalence transformations of the form
$X\mapsto S_1 X S_2$.
\end{proposition}
The proof of this statement is simple. We have just to exploit
that $S_i\sigma S_i^T=\sigma$ and that for any two matrices $AB$
and $BA$ have the same non-zero spectrum.

Note that the entire spectrum of $\Sigma(X)$ is two-fold
degenerate, i.e.,
$spec(\Sigma)=\{\lambda_1,\ldots,\lambda_n,\lambda_1,\ldots,\lambda_n\}$
\cite{root} and in addition complex eigenvalues come in conjugate
pairs $\lambda,\bar{\lambda}$
(as it holds for every real matrix).
 If $X$ is positive definite we recover, in fact, the normal mode
frequencies $\nu_k=\sqrt{\lambda_k}$ so that all invariants are
positive real numbers in this case. The occurrence of complex
eigenfrequencies is, in fact, a well known phenomenon in many
fields of physics in particular where harmonic approximations are
used in intermediate energy regimes. Examples can be found in
contexts from molecular condensates \cite{mol} to gravitational
waves \cite{gra} and sonic black holes \cite{son}.

While Prop.\ref{prop:inv} evidently holds for arbitrary
rectangular and possibly singular matrices $X$, we will for the
sake of simplicity restrict ourselves to non-singular square
matrices in the following.  Our aim is to show that
$\lambda_1,\ldots,\lambda_n$ are the only invariants and that they
essentially determine the normal-form of $X$ with respect to
symplectic equivalence transformations:

\begin{proposition}[Canonical form]\label{prop:main}
For every nonsingular matrix $X\in\mathbb{R}^{2n\times 2n}$ there
exist real symplectic transformations $S_1,S_2$ such that
\be\label{eq:normalform} S_1
X S_2=\left(%
\begin{array}{cc}
  \1_n & 0 \\
  0 & J \\
\end{array}%
\right),\ \ J=\left(%
\begin{array}{ccc}
  J_1(\lambda_1) & 0 & \ldots \\
  0 & J_2(\lambda_2) & \ldots \\
  \vdots & \vdots & \ddots \\
\end{array}%
\right),\ee where each $J_k$ is a real Jordan block \cite{Jblock}
corresponding to either a complex conjugate pair
$\{\lambda,\bar\lambda\}$ of eigenvalues of $\Sigma(X)$ or to one
of its real eigenvalues. In the former case the diagonal of
$J_k(\lambda)$ is build up out of real
$2\times 2$ blocks of the form \be \label{eq:2x2} \left(%
\begin{array}{cc}
  a & b \\
  -b & a \\
\end{array}%
\right),\quad \lambda=a+i b\;.\ee
\end{proposition}
\proof{First note that $\Sigma(X)$ is a skew-Hamiltonian matrix,
i.e., $(\Sigma\sigma)^T=-(\Sigma\sigma)$ (we drop the dependence
on $X$ in the following). For every real skew-Hamiltonian matrix
there exists a real symplectic similarity transformation such that
$S\Sigma S^{-1}=-(M\oplus M^T)$ is block diagonal \cite{root}.
Exploiting in addition that every real matrix $M$ can be written
as a product of two real symmetric
matrices $AB=M$ \cite{2sym} we can write \be\label{eq:W2} (SX)\sigma(SX)^T\sigma=\left(%
\begin{array}{cc}
  0 & A \\
  B & 0 \\
\end{array}%
\right)^2=:W^2\;.\ee
 Now define a transformation $S'$ by imposing $SXS'=W$. In fact, $S'$ is symplectic as can be seen from
 \be \label{eq6} S'^T\sigma S' = W^T \sigma^T \big[\sigma(SX)^{-T}\sigma(SX)^{-1} \big] W\\
 = \sigma, \ee
where we have used that $\sigma^T\sigma=\1$, $W^T\sigma^T=\sigma
W$ and that the expression in squared brackets is by
Eq.(\ref{eq:W2}) equal to $W^{-2}$.
$W^T\sigma^T=\sigma W$ is easily seen by exploiting the
$A,B$-block structure of $W$. Note that Eq.(\ref{eq6}) is the
point in proof where we use non-singularity of $X$.

To proceed we exploit the subgroup $GL(n)\subset Sp(2n)$, i.e.,
the fact that for every real invertible matrix $G$, the block
matrix $G^{-1}\oplus G^T$ is symplectic \cite{Mukunda}.
Multiplying $SXS'$ from the left with $A^{-1}\oplus A^T$ and from
the right with $\sigma$, which is a symplectic transformation in
its own right, we obtain $\1_n\oplus(-A^TB)$. This can be brought
to the claimed form in Eq.(\ref{eq:normalform}) via a symplectic
similarity transformation by $G^{-T}\oplus G$. Here we use the
real Jordan canonical form $J=G (-A^TB) G^{-1}$ in which complex
conjugate pairs of eigenvalues correspond to real $2\times 2$
matrices of the form in Eq.(\ref{eq:2x2}) (cf.\cite{HJ}). It
remains to show that the spectrum of $\Sigma(X)$ is a doubling of
the spectrum of $J$. By Prop.\ref{prop:inv} we have that
$spec(\Sigma(X))=spec(\Sigma(\1\oplus J))$ so that the identity
$\Sigma(\1\oplus J)=J^T\oplus J$ completes the proof.
 }

Some remarks on the normal form in Eq.(\ref{eq:normalform}) are in
order. First note that it is minimal in the sense that the number
of continuous parameters cannot be further reduced by symplectic
equivalence transformations as they are all invariants due to
Prop.\ref{prop:inv}. Similarly looking at the invariants tells us
that an entirely diagonal normal form, analogous to the usual
normal mode decomposition, cannot exist in general as real
diagonal matrices have only real invariants. Hence, there is no
way of diagonalizing the remaining $2\times 2$ blocks since they
correspond to complex $\lambda$'s. Concerning the possible
appearance of defective parts in the Jordan blocks \cite{Jblock}
we note that, as usual, they are not stable with respect to
perturbations. The fact that matrices with non-defective normal
form are dense can be seen by noting that for any
$X\in\mathbb{R}^{2n\times 2n}$ there exists an $X_\epsilon$
arbitrary close to it such that $spec(\Sigma(X_\epsilon))$ is only
two-fold degenerate (and non-singular). The corresponding $J$ has
no degeneracy and is thus non-defective.

The presented canonical from can be regarded as a generalization
of the seminal results by Williamson \cite{Williamson} and its
extensions \cite{canonicalforms} on normal forms of symmetric (not
necessarily positive) matrices under symplectic transformations.
For positive definite matrices $X$ the canonical form in
Eq.(\ref{eq:normalform}) and the usual normal mode decomposition
in Eq.(\ref{eq:Williamson}) coincide up to a simple squeezing
transformation and the invariants are related via
$\lambda_k=\nu_k^2$.

In the following we will discuss two applications of the above
result in contexts where the two sets of modes on which $S_1$ and
$S_2$ act on either correspond to two different parties (Alice and
Bob, say) or to input and output of a quantum channel. Note that
in both cases the two sets need not be of the same physical type.
A prominent example of that form is a set of light modes coupled
to modes of collective spin fluctuations in atomic ensembles
\cite{Klemens}.

\begin{figure}[t]
\begin{center}
\epsfig{file=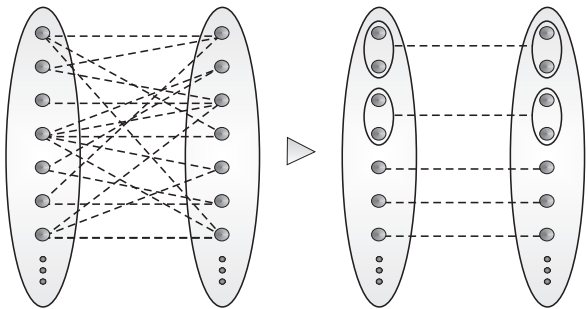,angle=0,width=0.9\linewidth}
\end{center}
\caption {{\it Left}: let Alice and Bob each have a collection of
oscillators with arbitrary correlations between them (indicated by
dashed lines). {\it Right}: there exist local canonical
transformations which condense the correlations into elementary
units (single modes or pairs) and eliminate all
cross-correlations.} \label{fig1}
\end{figure}

\section{Decoupling and Condensing Correlations}

Consider a bipartitioned collection of $n+n$ modes as in Fig.1.
The covariance matrix $\Gamma\in\mathbb{R}^{4n\times4n}$
can then be partitioned into blocks\be \Gamma=\left(%
\begin{array}{cc}
  \Gamma_A & X \\
  X^T & \Gamma_B \\
\end{array}%
\right),\ee where $\Gamma_A,\;\Gamma_B$ are the local covariance
matrices and $X$ describes correlations between the two parts. A
local symplectic transformation $\Gamma\mapsto (S_A\oplus
S_B)\Gamma(S_A\oplus S_B)^T$ transforms the correlation block as
$X\mapsto S_AXS_B^T$. Hence, Prop.\ref{prop:main} can be directly
applied to decouple and condense the correlations. In the generic
case of non-singular, non-defective $X$ we are then left with
correlated single modes whose correlations are characterized by
real invariants $\lambda$ and with correlated pairs corresponding
to complex $\lambda$ with a correlation block of the form in
Eq.(\ref{eq:2x2}). All the cross-covariances between opposite
subsets which do not correspond vanish (see Fig.1).

A particular known instance of this result is the case of pure
Gaussian quantum states. For pure bipartite states correlations,
which are then due to entanglement, and local spectral properties
 determine each other. This is the content of the Schmidt
decomposition which in terms of the covariance matrix implies that
$\Gamma_A,\;\Gamma_B$ and $X$ can then be simultaneously
diagonalized by local symplectic transformations such that
$\nu_k=\sqrt{1-\lambda_k}$ ($\lambda_k\leq 0$) is the mean
particle number in the $k$'th normal mode of each site
\cite{Schmidt,HW}.

This pure state normal form simplified investigations in various
directions like security proofs in quantum cryptography \cite{QKD}
or the transformation \cite{tent}, localization \cite{lent} and
characterization \cite{cent} of entanglement.
Prop.\ref{prop:main}. now provides the analogous normal form for
mixed states, which again can be regarded as a condensation of
correlations. As such, it might be a useful first step in quantum
information protocols which use correlations as a resource.

\begin{figure}[ttt]
\begin{center}
\epsfig{file=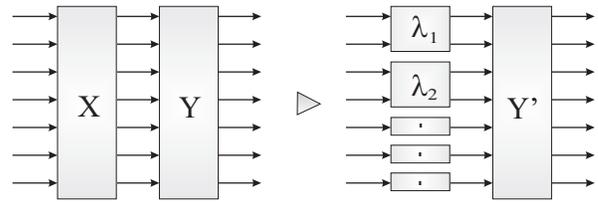,angle=0,width=0.9\linewidth}
\end{center}
\caption {{\it Left}: A multi-mode Gaussian channel consists out
of an interaction part $X$, which directly couples the input
modes, and an addition of input-independent noise $Y$. {\it
Right}: The interaction part is decoupled into single-mode and
two-mode parts by applying canonical transformations before and
after the channel.  Two-mode interactions corresponding to a
complex invariant $\lambda$ only occur if the global
system-plus-environment evolution involves squeezing.}
\label{fig1}
\end{figure}

\section{Decoupling of Interactions}

Let us now consider dissipative evolutions of multi-mode quantum
systems. An important class of such evolutions are those where
system plus environment undergo a global canonical transformation.
For the system (with traced out environment) this leads to
so-called Gaussian or quasi-free channels \cite{HW,EW}
realized by optical fibres or, if the transformation is in time
rather than in space, by quantum memories build upon atomic
ensembles \cite{Qmemory}.

Gaussian channels are characterized by a pair of matrices
$X,Y\in\mathbb{R}^{2n\times2n}$ satisfying the constraint $i
X^T\sigma X+Y\geq i\sigma$. The covariance matrix evolves then
according to \be \Gamma\mapsto X^T\Gamma
X+Y.\label{eq:Gchannel}\ee That is, $X$ can be regarded as
characterizing direct interactions between the modes and $Y$ is a
noise-term which is input-independent. If we now apply a
symplectic transformation before and after the channel then
$X,Y\mapsto S_1XS_2,S_2^TYS_2$. We can thus again exploit
Prop.\ref{prop:main} in order to simplify the structure of the
interaction part of the evolution. This way of encoding and
decoding information sent through the channel has been
successfully exploited in the context of channel capacities of
single-mode channels \cite{capacities} for which it leads to a
simple normal form \cite{1normalform}. For multi-mode channels it
allows us, in the generic case, to reduce $X$ to two-mode
interactions of the form in Eq.(\ref{eq:2x2}) and single-mode
parts (see Fig.2).

The appearance of pair interactions corresponding to complex
invariants $\lambda$ is, in fact, a signature of a non-number
preserving system-environment interaction. In order to see this
recall that a global number-preserving transformation can be
written as \be S=\left(%
\begin{array}{cc}
  C & D \\
  -D & C \\
\end{array}%
\right),\label{eq:U}\ee where $C+iD$ is a unitary and the block
structure in Eq.(\ref{eq:U}) refers to a decomposition of phase
space into position and momentum space (rather than system and
environment). For the reduced system evolution this leads to an
$X$ which has the same structure as $S$ but without the
restriction of the matrices being real and imaginary parts of a
unitary. Let us denote the corresponding blocks in $X$ by $c$ and
$d$ and calculate \be \Sigma(X)=\left(%
\begin{array}{cc}
  dd^T+cc^T & dc^T-cd^T \\
  cd^T-dc^T & dd^T+cc^T \\
\end{array}%
\right).\ee As $\Sigma(X)$ is Hermitian it has indeed only real
eigenvalues $\lambda_k$ which shows that pair interactions in the
normal form which correspond to complex $\lambda$'s witness a
squeezing-type interaction between system and environment.

\section*{Acknowledgments} The author thanks A.S. Holevo for many
inspiring discussions on the topic.


\begin{thebibliography}{99}
\bibitem{Jblock} Each real Jordan block has the form
$$ J(\lambda)=\left(%
\begin{array}{cccc}
  \Lambda & \1_2 &  &  \\
   & \Lambda & \ddots &  \\
   &  & \ddots & \1_2 \\
   &  &  & \Lambda \\
\end{array}%
\right),\ \ \Lambda=\left(%
\begin{array}{cc}
  a & b \\
  -b & a \\
\end{array}%
\right),\ \ \lambda=a+ib, $$ and it is called non-defective if
$J(\lambda)=\Lambda$.
\end{thebibliography}
\end{document}